\documentclass[conference]{IEEEtran}
\usepackage{graphicx,amsmath,amssymb,cite,algorithm,algorithmic}
 
\begin{document}
\IEEEoverridecommandlockouts
\title{Bit Error Rate Analysis\\ of Cooperative Beamforming for Transmitting\\ Individual Data Streams}
\author{
\IEEEauthorblockN{Spyridon Vassilaras\IEEEauthorrefmark{1}, George C. Alexandropoulos\IEEEauthorrefmark{1}, and Antonis A. Kalis\IEEEauthorrefmark{2}}
\IEEEauthorblockA{\IEEEauthorrefmark{1}Mathematical and Algorithmic Sciences Lab, France Research Center, Huawei Technologies Co. Ltd., Paris, France}
\IEEEauthorblockA{\IEEEauthorrefmark{2}SignalGeneriX Ltd., Limassol, Cyprus}
e-mails: \{spyros.vassilaras, george.alexandropoulos\}@huawei.com, akal@signalgenerix.com
\thanks{Part of this work was performed while the first two authors were with the Broadband Wireless and Sensor Networks Group, Athens Information Technology, Peania, Greece.}
}
\maketitle
\begin{abstract}
Cooperative beamforming (CB) has been proposed as a special case of coordinated multi-point techniques in wireless communications. In wireless sensor networks, CB can enable low power communication by allowing a collection of sensor nodes to transmit data simultaneously to a distant fusion center in one hop. Besides the traditional CB approach where all nodes need to share and transmit the same data, a more recent technique allows each node to transmit its own data while still achieving the benefits of cooperation. However, the intricacies of varying beamforming gains in the direct sequence spread spectrum with binary frequency shift keying multiple access scheme used in this context need to be taken into account when evaluating the performance of this beamforming technique. In this paper, we take the first step towards a more comprehensive understanding of this individual-data CB technique by proposing a best suited decoding scheme and analyzing its bit error rate (BER) performance over an additive white Gaussian noise channel. Through analytical expressions and simulation results BER curves are drawn and the achieved performance improvement offered by the CB gain is quantified.
\end{abstract}

\section{Introduction}
In recent years, there has been increasing interest in Coordinated Multi-Point (CoMP) transmission in which a set of independent transmitters are forming a virtual antenna array so as to achieve multi-antenna communication advantages \cite{cooperative_gesbert}. Cooperative Beamforming (CB) is a special case of CoMP techniques in which a collection of transmitters coordinate their transmissions in order to achieve directional antenna patterns in the far-field. In this paper we focus on the use of CB in Wireless Sensor Networks (WSNs) that are deployed in a large open field. 

WSNs are becoming important in a number of scientific areas, including environmental monitoring, structural health monitoring and security surveillance \cite{Survey_WSNs}. Wireless sensors are usually small in physical dimensions and operated by battery power. After the deployment of sensors over a large area, their batteries replacement or recharging can be a tedious and costly task. Hence extending the energy lifetime of WSN is extremely important. In addition, information transmission in WSN is commonly accomplished through the use of multi-hop communication \cite{Royer_review, Ephremides, Stojmenovic}, according to which nodes play the role of relays to convey information to a distant fusion center where this information will be stored and processed. But this kind of communication scheme requires multiple transmissions based on WSN-suitable Medium Access Control (MAC) and routing protocols, which translates to end-to-end transmission delays and high energy consumption. CB has been proposed as a means to bypass multi-hop transmissions, so that nodes can access the fusion center directly, thus achieving a one-hop connection \cite{cooperative_beamforming, collaborative_poor, Lun_CB, Ahmed_CB, Palomar_2014}.

In conventional CB, all cooperating nodes transmit the same data such that the transmitted signals in the desired direction are coherently added. However, the disadvantage with this approach is that the information has to be shared among cooperating nodes prior to their one-hop transmission to the distant fusion center, burdening the network with an initial data broadcasting phase which brings all the shortcomings of multi-hop communication mentioned above. In a variation of this scheme, nodes in a WSN are divided into clusters and all nodes in a given cluster are cooperating to perform one-hop CB communication. In \cite{collaborative_poor}, the performance of same-data CB is analyzed using the theory of random arrays \cite{Lo_antenna_arrays}.

In \cite{cooperative_kalis}, the authors proposed a novel scheme using CB and Spread Spectrum (SS) techniques to overcome the limitation in \cite{collaborative_poor} and allow each node to transmit its own data. More specifi\-cally, Direct Sequence SS (DSSS) with Binary Frequency Shift Keying (BFSK) modulation was employed in order to achieve multiple channel access. Under this scheme while each node transmits its own data, at any point in time it transmits either an analog sinusoidal wave at frequency $f_1$ or a similar wave at frequency $f_2$. Furthermore the sinusoidal waves of the same frequency from all nodes are co-phased and thus all sinusoids of frequency $f_1$ add up at the Receiver (RX) (and the same holds for all sinusoids of frequency $f_2$). In \cite{onehop_kalis}, the directive beamforming gain achieved by this individual-data CB technique is calculated for different sensor network densities. The authors argue that, based on the law of large numbers and the fact that each node transmits at $f_1$ or $f_2$ with equal probability, about half of the nodes will be transmitting at either frequency at any point in time. Although this is a fair assumption in order to calculate the expected beamforming gain of the individual-data CB technique, a more comprehensive investigation of the beamforming gain achieved at each point in time needs to be performed in order to obtain the bit error rate (BER) performance of this technique. In this paper, we propose a decoding scheme for the individual-data CB technique and evaluate its BER performance over an Additive White Gaussian Noise (AWGN) channel as a function of the number of nodes, the Signal-to-Noise Ratio (SNR) and the distance from the deployment area to the fusion center. Such BER vs SNR curves allow for determining the required transmit power for a desired BER and thus energy budgeting and energy consumption evaluation of the considered individual-data CB technique.

\section{System Model}\label{Sec:Sys_Model}
\begin{figure}[t!] 
\centering
\includegraphics[width=7.2cm,height=5.25cm]{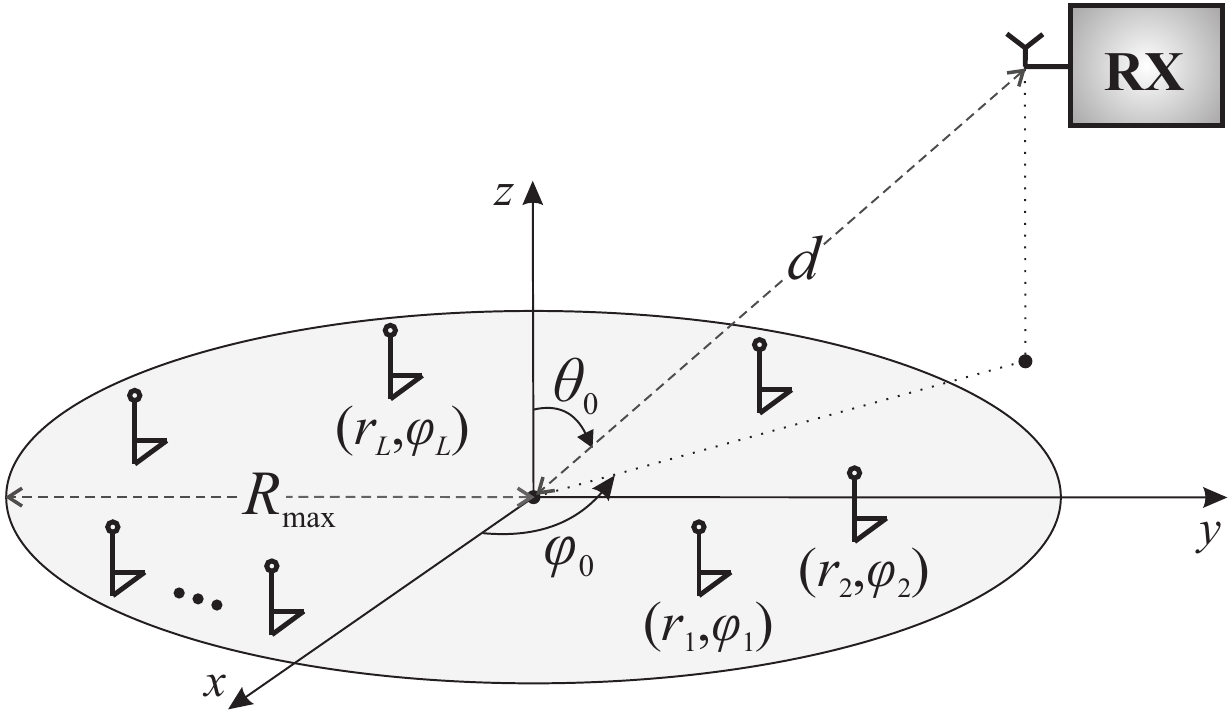}
\caption{The considered wireless sensor network topology.}
\label{network_topology}
\end{figure}

WSNs are characterized by the dense deployment of sensor nodes that continuously observe physical phenomena with applications in a wide range of fields. In this paper, we assume that a number of sensors need to be deployed in an open field in order to collect real-time data. As all nodes are measuring the same environmental parameters at equal sampling rates, the generated data streams are expected to be of equal source data rate. The field is assumed to be close to planar, i$.$e$.$, elevation differences are negligible. We further assume that the sensor nodes are uniformly distributed in a circular area with radius $R_{\max}$ as shown in  Fig.~\ref{network_topology}. The common RX which collects data from all nodes (fusion center) is located far away from the sensor area, i$.$e$.$, it is assumed that the distance of RX from the center of the sensor area is $d\gg R_{\max}$. Each node is equipped with a single isotropic antenna.
The location of RX in spherical coordinates is assumed to be $(d,\theta_0,\varphi_0)$, where $\theta_0\in[0,\pi]$ is the elevation angle and $\varphi_0\in[-\pi, \pi]$ is the azimuth angle. The location of each of the $L$ sensors on the plane can be represented by $(0,r_l,\varphi_l)$\footnote{Unless otherwise stated, for indices $l$ and $i$, which will be introduced later on, holds $l=1,2,\ldots,L$ and $i=1,2$.}, where $r_l\in [0,R_{\max}]$ and $\varphi_l\in[-\pi, \pi]$. Also, the density of the sensor nodes within the area of $\pi R_{\max}^2$ is low enough so that the inter-sensor distances are sufficiently large in order for unwanted mutual coupling among the sensor nodes' antennas to be avoided. 

\begin{figure}[t!]
\centering
\includegraphics[width=8.8cm,height=5.45cm]{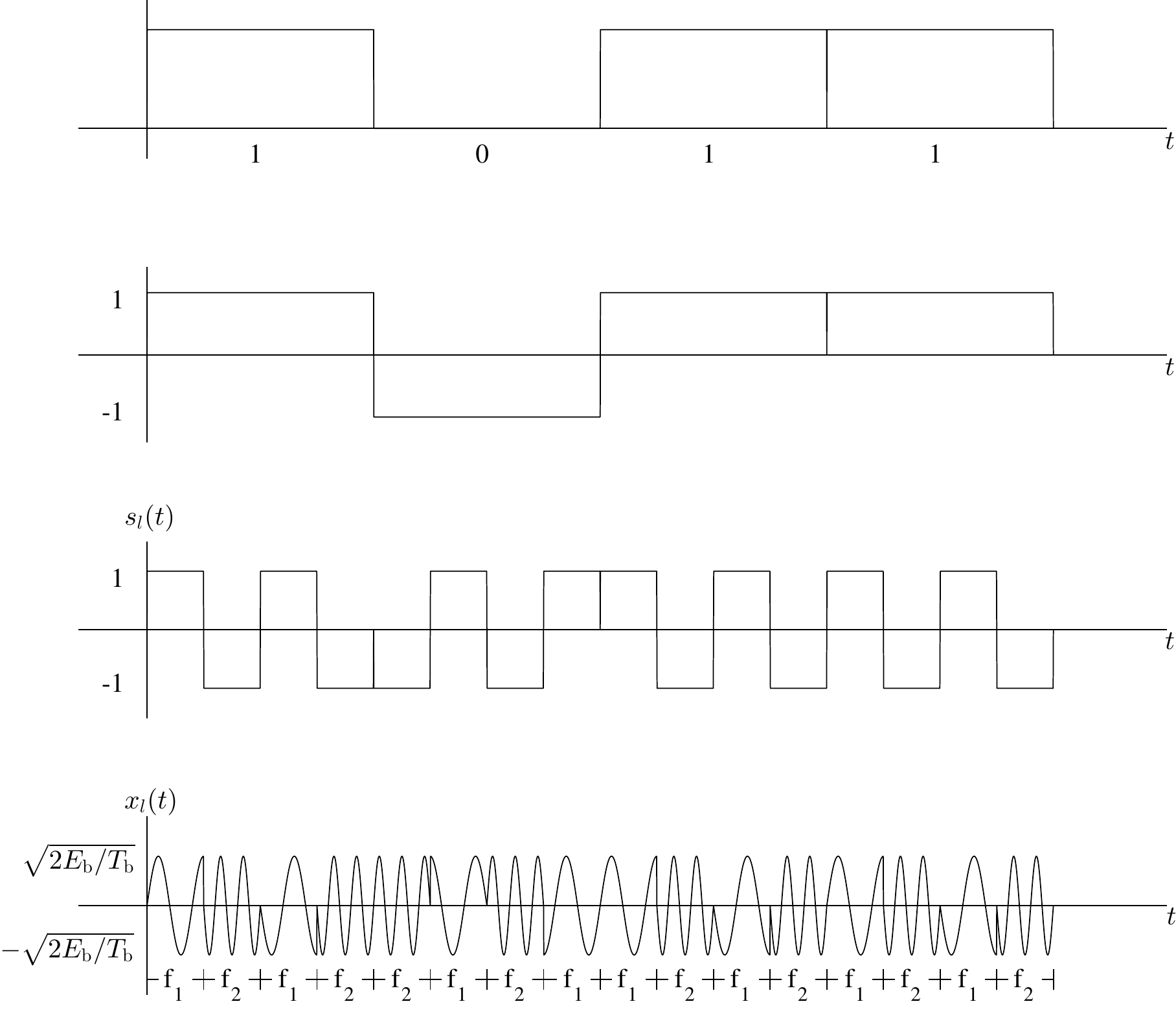}
\caption{Toy example of a sensor node using the spreading code $[1,-1,1,-1]$ and transmitting the series of bits $1,0,1,1$ (top plot). The series of bits is first written as $1,-1,1,1$ (2\textsuperscript{nd} plot) and then each bit is spread by the spreading code to obtain the chip sequence (3\textsuperscript{rd} plot). Finally, the resulting chip sequence is modulated to obtain the signal to be transmitted (bottom plot).}
\label{dsss_figure}
\end{figure}

The mechanism of the considered CB scheme is based on the following principle: Consider a beacon station, placed at a symmetric position to the intended RX with respect to the $z$-axis, which sends a sinusoidal synchronization signal to all sensor nodes. The sinusoidal beacon signal is received at each node with a slight phase difference due to different propagation delays. If all sensor nodes rebroadcast this beacon signal with an identical rebroadcast delay, it has been shown that a strong beam is produced in the direction of RX \cite{cooperative_kalis}. Furthermore, at the exact RX point, all sinusoidal signals sent by the various sensors are co-phased since the total distance beacon station - sensor node - RX is the same for all nodes. Note that this method of synchronization based on an external beacon signal alleviates the need of using accurately synchronized internal clocks for achieving the same beamforming result. Clearly, if all nodes use the sinusoidal signal to modulate the exact same data stream, the beamforming property is maintained. But as already discussed, using same-data CB to collect and transmit each node's individual data streams has several drawbacks. On the other hand, individual-data CB has been proposed in \cite{cooperative_kalis} based on the following simple and elegant idea. By using a DSSS/BFSK channel access and modulation scheme, each node's individual transmission consists of a series of short sinusoidal waveforms with frequency either $f_1$ or $f_2$. Therefore, although each node transmits a different sequence of waveforms, at any point in time a set of nodes transmits the same $f_1$ waveform and the remaining nodes the same $f_2$ waveform.      

DSSS/FSK hasn't received much attention in the literature due to the clear implementation advantages of the DSSS Phase-Shift Keying (DSSS/PSK) and frequency-hopping SS FSK schemes in cellular telephony and wireless local area network applications. DSSS/PSK is well studied for single- and multi-user scenarios \cite{book_hanzo}. PSK however cannot be used in our specific beamforming application since its random phase changes prevent co-phased signal transmission which is necessary in order to achieve CB. A version of DSSS with FSK in which the digital bit stream is first modulated using FSK and then spread (by multiplying the modulated signal by the spreading mask) which is studied in \cite{Sunkara_thesis} is also not suitable for beamforming (for the same reason).     

In order to achieve CB with individual data per sensor, the digital bit stream of each sensor needs to be first spread using Walsh-Hadamard orthogonal spreading codes \cite{book_hanzo} and then BFSK modulated by transmitting a ``$1$'' chip using $f_1$ and a ``$-1$'' chip using $f_2$. This is illustrated in Fig$.$~\ref{dsss_figure}. In this paper we also consider coherent demodulation, which is best suited for wireless communication when all transmitters and RXs are fixed, and therefore the signal phase at RX is assumed to be known. To ensure orthogonality for our coherent BFSK system, the minimum separation between the frequency tones should be equal to the chip rate $R_{\rm ch}$. Hence, the transmitted bandpass signal from the $l$-th sensor node at each time interval $0\leq t\leq T_{\rm ch}$ ($T_{\rm ch}=1/R_{\rm ch}$ is the chip period) can be mathematically expressed as
\begin{equation}\label{Eq:TX_signal}
x_l(t) = \sqrt{\frac{2E_{\rm b}}{T_{\rm b}}}\cos\left\{2\pi\left[f_c + \frac{s_l\left(t\right)}{2T_{\rm ch}}\right]t+\phi_i\right\}
\end{equation}
where $E_{\rm b}$ is the transmission energy per bit, $T_{\rm b}$ is the bit duration and $f_c$ is the carrier frequency. Furthermore, $s_l(t)$ denotes the complex envelope of the DSSS signal for the $l$-th sensor node and is given by
\begin{equation}\label{Eq:Complex_Envelope}
s_l(t) = \sum_n\sum_{k=0}^{L-1} b_n^{(l)}c_k^{(l)}p\left[t-(n-1)T_{\rm b}-kT_{\rm ch}\right]
\end{equation}
where $b_n^{(l)}=\pm1$ is the $n$-th ($n=1,2,\ldots$) information bit of the $l$-th sensor node and $c_k^{(l)}=\pm1$ denotes the $k$-th ($k=0,1,\ldots,L-1$) bit of its spreading code sequence of length $L$; this code lenght ensures the orthogonality of the spreading code sequences used. In addition, $p\left[t-(n-1)T_{\rm b}-kT_{\rm ch}\right]$ in \eqref{Eq:Complex_Envelope} represents rectangular pulses with unit amplitude in the time interval $(n-1)T_{\rm b}+kT_{\rm ch}\leq t\leq (n-1)T_{\rm b}+(k+1)T_{\rm ch}$ and $0$ elsewhere. Finally, $\phi_i$ in \eqref{Eq:TX_signal} is the phase introduced by the BFSK modulator that depends on the two frequency components:
\begin{equation}\label{Eq:Frequencies}
f_{1,2} = f_c\pm\frac{1}{2T_{\rm ch}}.
\end{equation}
From the theory underlining DSSS with orthogonal codes, it is known that digital decoding is achieved by multiplying the received signal by the individual mask of each transmitter, since it holds
\begin{equation}\label{Eq:digital_despreading}
\left\{\sum_{k=1}^{L} b_n^{(k)} \left[c_0^{(k)}\,c_1^{(k)}\,\cdots\,c_{L-1}^{(k)}\right] \right\}  \left[c_0^{(l)}\,c_1^{(l)}\,\cdots\,c_{L-1}^{(l)}\right]^{\rm T} = L b_n^{(l)}
\end{equation}
with the symbol ${\rm T}$ denoting vector transposition.

Unlike traditional DSSS systems where calculating the aggregate received signal is straightforward, in the considered system we need to take into account the varying beamforming gain due to the varying number of superimposed co-phased sinusoidal signals of frequency either $f_1$ or $f_2$. Suppose that at each point $t$ in time a subset $\mathcal{L}_{t,1}$ of all nodes transmits at $f_1$ and the remaining nodes in subset $\mathcal{L}_{t,2}$ transmit at $f_2$, where $\forall\,t$, $|\mathcal{L}_{t,1}| + |\mathcal{L}_{t,2}| = L$ ($|\mathcal{A}|$ denotes the cardinality of a set $\mathcal{A}$). Let also the beamforming power gain of subsets $\mathcal{L}_{t,1}$ and $\mathcal{L}_{t,2}$ at a specific $t$ be given by $G\left(\mathcal{L}_{t,1}\right)$ and $G\left(\mathcal{L}_{t,2}\right)$, respectively. Then, the received bandpass signal at RX can be expressed as
\begin{equation}\label{Eq:RX_signal}
y(t) = \sqrt{\frac{2E_{\rm b}}{T_{\rm b}}P(d)}\sum_{i=1}^2A\left(\mathcal{L}_{t,i}\right)\cos\left(2\pi f_it+\phi_i\right)+n(t)
\end{equation}
where $P(d)$ denotes the free-space path loss that depends on the distance to RX (the dependence of $P(d)$ on $f_i$'s is assumed to be negligible) and the amplification $A\left(\mathcal{L}_{t,i}\right)$ is the product of the square root of the beamforming gain times the number of transmitting nodes: 
\begin{equation}\label{Eq:A_parameters}
A\left(\mathcal{L}_{t,i}\right) = \sqrt{G\left(\mathcal{L}_{t,i}\right)}\left|\mathcal{L}_{t,i}\right|.
\end{equation}
In addition, $n(t)$ in \eqref{Eq:RX_signal} represents the zero-mean complex AWGN with variance $\sigma_N^2$. It is noted that $G\left(\mathcal{L}_{t,i}\right)$ depends not only on the number of nodes transmitting at frequency $f_i$ but also on the exact locations of these nodes. 

\section{Cooperative Beamforming Gain}\label{Sec:Gain}
In order to calculate the beamforming gain, we adopt all the assumptions of \cite[Sec$.$~II]{collaborative_poor} which are reasonable in our open-space WSN scenario. We also consider the case (without loss of generality) where the azimuth angle of the common RX is $\phi_0 = 0$.  Then, a closed form expression for the CB directivity can be derived based on the analysis in \cite{collaborative_poor} as
\begin{equation}\label{eq_directivity}
D(\mathcal{L}_{t,i},\theta_0) = \frac{2\pi}{\left|\mathcal{L}_i\right|^2}\left(\int_{-\pi}^\pi\left|\sum_{l\in\mathcal{L}_{t,i}} e^{-j\alpha\left(\theta_0,\varphi\right)z_l\left(\varphi\right)}\right|^2{\rm d}\varphi\right)^{-1}
\end{equation}
where the function $\alpha\left(\theta_0,\varphi\right)$ is given by
\begin{equation}\label{alpha}
\alpha\left(\theta_0,\varphi\right) = 4\pi\frac{R_{\max}}{\lambda_i}\sin\left(\theta_0\right)\sin\left(\frac{\varphi}{2}\right).
\end{equation}
Furthermore, the function $z_l\left(\varphi\right)$ depends on the azimuth angle $\varphi_l$ of the $l$-th node and is defined as
\begin{equation}\label{z}
z_l\left(\varphi\right) = \frac{r_l}{R_{\max}}\sin\left(\varphi_l-\frac{\varphi}{2}\right).
\end{equation}
In \eqref{alpha}, the quantity $R_{\max}/\lambda_i$ is referred to as the radius normalized to the wavelength $\lambda_i= c / f_i$ with $c$ being the speed of light. Throughout this paper, we assume that RX is placed at an elevation angle of $\theta_0=\pi/3$, and hence dependency on $\theta_0$ is dropped. The beamforming gain is obtained from the directivity in \eqref{eq_directivity} as $G\left(\mathcal{L}_{t,i}\right) = \rho D\left(\mathcal{L}_{t,i},\pi/3\right)$, where $\rho$ is the efficiency factor assumed to be equal to unity.
\begin{figure}[t!]
\centering
\includegraphics[keepaspectratio,width=3in]{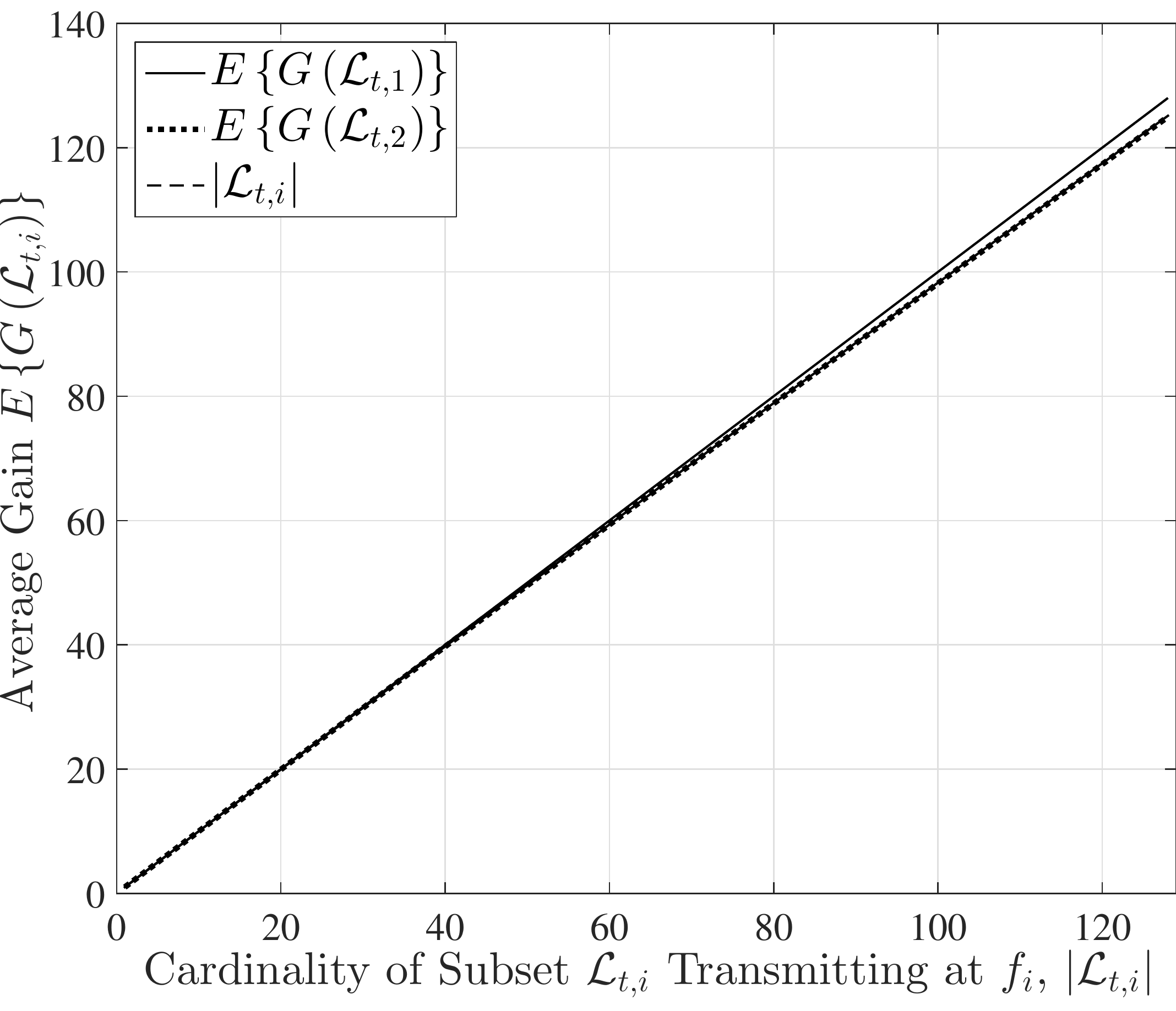}
\caption{The average beamforming gain $G\left(\mathcal{L}_{t,i}\right)$ $\forall$ $i$ is approximately equal to the number of transmitting sensor nodes, $\left|\mathcal{L}_{t,i}\right|$.}
\label{gain_vs_num_nodes}
\end{figure}

\begin{figure}[t!]
\centering
\includegraphics[keepaspectratio,width=3in]{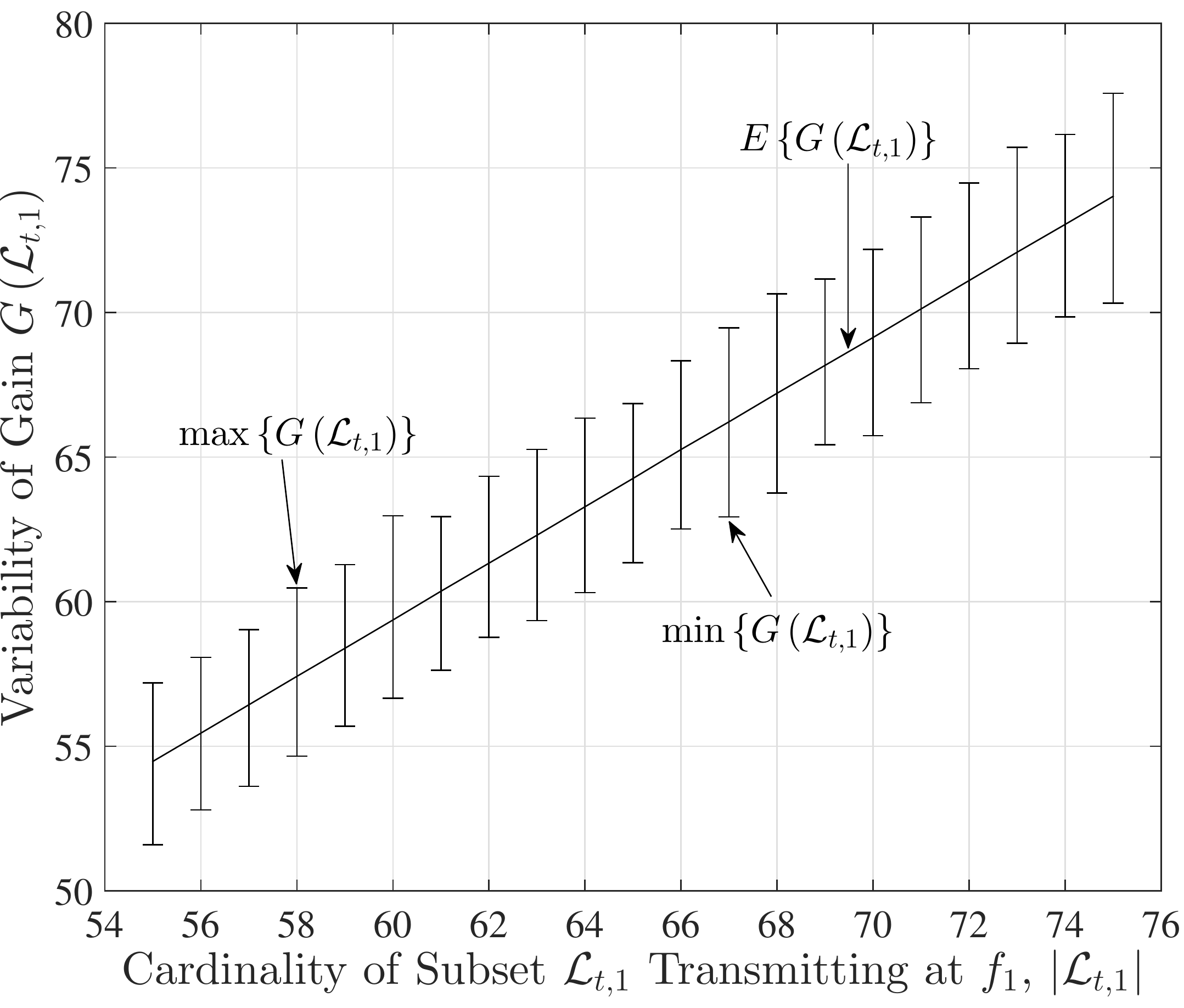}
\caption{Minimum, average and maximum values of the beamforming gain $G\left(\mathcal{L}_{t,1}\right)$ for different numbers of transmitting sensor nodes, $\left|\mathcal{L}_{t,1}\right|$.}
\label{min_max_gain}
\end{figure}

In order to investigate the variability of the achieved power gain $G\left(\mathcal{L}_{t,i}\right)$ as a function of the number and the particular subset of sensor nodes transmitting at the same frequency, we performed the following simulation using \textsc{Matlab}. First we placed a set of $L=128$ nodes at random positions uniformly distributed in a circular area with $R_{\max} = 100$ m. Then for every $|\mathcal{L}_{t,i}|\in\{1,2,\ldots,L\}$ we randomly selected $10^4$ subsets $\mathcal{L}_{t,i}$ of these same $128$ nodes. For each subset we calculated $G\left(\mathcal{L}_{t,i}\right)$ and then computed $E\{G\left(\mathcal{L}_{t,i}\right)\}$ with $E\{\cdot\}$ denoting the expectation operator, $\min\{G\left(\mathcal{L}_{t,i}\right)\}$ and $\max\{G\left(\mathcal{L}_{t,i}\right)\}$ of these $10^4$ values of the gain. In this paper, we assumed $f_{1,2}=2.45$ GHz $\pm64$ KHz. 
This frequency separation ensures orthogonality of the two frequency tones for $R_{\rm ch}=128000$ chips/sec corresponding to a bit rate of $1$ Kbps per node. For this small difference in the two frequencies the obtained beamforming gains for $f_1$ and $f_2$ are almost identical. The first important observation from this experiment is that the average gain is approximately equal to the number of transmitting nodes, i$.$e$.$, $E\{G\left(\mathcal{L}_{t,i}\right)\}\cong|\mathcal{L}_{t,i}|$ $\forall\,t$. This is illustrated in Fig$.$~\ref{gain_vs_num_nodes} which shows an almost perfect fit of the values of the average gain on the $y = x$ line. The second important observation is that the intervals of possible values of the gain for adjacent number of nodes are well overlapping. This means that if RX wants to determine the number of transmitting nodes, the probability of decoding error will be substantial even without any AWGN. In other words, the gain variability due to different equal-size subsets of nodes transmitting cannot be neglected. This finding is illustrated in Fig$.$~\ref{min_max_gain} where we depict $\min\{G\left(\mathcal{L}_{t,1}\right)\}$, $E\{G\left(\mathcal{L}_{t,1}\right)\}$ and $\max\{G\left(\mathcal{L}_{t,1}\right)\}$ values for $|\mathcal{L}_{t,1}|$ varying from $55$ to $75$. In the following section we will describe how these two findings affect the BER performance of the individual-data CB technique.   

\section{BER Analysis of Individual-Data CB\\ over AWGN Channels}\label{Sec:Analysis}
In this section we analyze the performance of the individual-data CB technique \cite{cooperative_kalis} which utilizes DSSS/BFSK as presented in Section~\ref{Sec:Sys_Model}. In particular, the decision statistics of a bit level decoding scheme over a AWGN channel are first derived and then a closed-form analytical approximate expression for its BER performance  is presented. 

\subsection{Decision Statistics}\label{Sec:Decision_Statistics}
The RX architecture of the considered individual-data CB technique \cite[Fig$.$~12]{cooperative_kalis} comprises of: \textit{i}) A BFSK demodulator operating at the chip level and aiming at computing the beamforming gains of the down-converted transmitted signals at frequencies $f_{1,2}'=\pm1/(2T_{\rm ch})$, that is \textit{ii}) followed by a bank of $L$ parallel decoders responsible for despreading each node's information bit stream. Assuming perfect knowledge of $\phi_i$'s at RX, the two outputs of the BFSK demodulator at the $q$-th ($q=1,2,\ldots$) chip level are given by
\begin{equation}\label{Eq:Decisions_BFSK}
\zeta_{q,i} = \frac{\sqrt{2T_{\rm b}}}{T_{\rm ch}}\int_{(q-1)T_{\rm ch}}^{qT_{\rm ch}}y(t)\cos\left(2\pi f_i't\right){\rm d}t
\end{equation}
and are then subtracted to obtain the (noisy) equivalent of the received aggregate chip sequence (to be explained in the sequel). More specifically, substituting \eqref{Eq:A_parameters} into \eqref{Eq:RX_signal}, then setting $\phi_i$'s equal to $0$ and substituting the result into \eqref{Eq:Decisions_BFSK}, and finally performing the subtraction $\kappa_q \triangleq \zeta_{q,1}-\zeta_{q,2}$ yields
\begin{equation}\label{Eq:Decision_BFSK}
\kappa_q = \sqrt{E_{\rm b}P(d)}\left[\sqrt{G\left(\mathcal{L}_{q,1}\right)}\left|\mathcal{L}_{q,1}\right|-\sqrt{G\left(\mathcal{L}_{q,2}\right)}\left|\mathcal{L}_{q,2}\right|\right]+w_q
\end{equation}
where $\mathcal{L}_{q,i}$ denotes the subset of nodes transmitting at $f_i$ in the $q$-th chip's time interval $(q-1)T_{\rm ch}\leq t\leq qT_{\rm ch}$ and random variable (RV) $w_q$ is obtained as
\begin{equation}\label{Eq:Diff_Noise}
\begin{split}
w_q = \frac{\sqrt{2T_{\rm b}}}{T_{\rm ch}}\int_{(q-1)T_{\rm ch}}^{qT_{\rm ch}}n(t)\left[\cos\left(2\pi f_1't\right)-\cos\left(2\pi f_2't\right)\right]{\rm d}t.
\end{split}
\end{equation}
It is noted that, using a similar analysis to \cite[Sec$.$~5$.$1]{B:Proakis_Dig_Com} for the projection of $n(t)$ onto the two sinusoidal basic functions, it can be shown that $w_q$ is a zero-mean Gaussian RV with variance $2\sigma_N^2T_{\rm b}/T_{\rm ch}$.
Also note that if the beamforming gains were all equal to $1$ then we would have:
\begin{equation}\label{Eq:noisy_digital_rx}
\kappa_q^\prime = \sqrt{E_{\rm b}P(d)}\left(\left|\mathcal{L}_{q,1}\right|-\left|\mathcal{L}_{q,2}\right|\right)+w_q
\end{equation}
and that the quantity $\left|\mathcal{L}_{q,1}\right|-\left|\mathcal{L}_{q,2}\right|$ is the transmitted aggregate chip value (at chip $q$). This is why $\kappa_q$ can be considered as the noisy and perturbed (by the different beamforming gains) equivalent of the received aggregate chip sequence.

Now there are two ways of proceeding in decoding the received $\kappa_q$'s in a bit interval: \textit{i}) \textit{Chip level decoding} where the received aggregate chip sequence is estimated at each chip and then multiplied by each node's spreading code sequence; and \textit{ii}) \textit{Bit level decoding} where the signal is first multiplied by the spreading code and then decoded. Our simulation results revealed that bit level decoding offers a better performance and therefore we will focus on this decoding scheme. In bit level decoding, by grouping each bit's $L$ $\kappa_q$'s into a vector and then taking the inner product with each node's spreading code sequence, the bank of $L$ parallel decoders outputs the decision statistics of each node's information bit stream. To this end, the decision statistics for the $n$-th information bit of the $l$-th sensor node is derived as
\begin{equation}\label{Eq:Decision_Stats_final_definition}
\hat{b}_n^{(l)} = \left[\kappa_{(n-1)L+1}\,\kappa_{(n-1)L+2}\,\cdots\,\kappa_{nL}\right]\left[c_0^{(l)}\,c_1^{(l)}\,\cdots\,c_{L-1}^{(l)}\right]^{\rm T}.
\end{equation}
By substituting \eqref{Eq:Decision_BFSK} into \eqref{Eq:Decision_Stats_final_definition}, $\hat{b}_n^{(l)}$ is obtained as
\begin{equation}\label{Eq:Decision_Stats_final}
\begin{split}
\hat{b}_n^{(l)} =& \sqrt{E_{\rm b}P(d)}\sum_{m=1}^{L}\left\{\sqrt{G\left[\mathcal{L}_{(n-1)L+m,1}\right]}\left|\mathcal{L}_{(n-1)L+m,1}\right|\right.
\\&\left.-\sqrt{G\left[\mathcal{L}_{(n-1)L+m,2}\right]}\left|\mathcal{L}_{(n-1)L+m,2}\right|\right\}c_{m-1}^{(l)}+\bar{w}_n
\end{split}
\end{equation}
where $\bar{w}_n$ is a Gaussian distributed RV given by
\begin{equation}\label{Eq:Sum_of_noise_terms}
\bar{w}_n=\sum_{m=1}^{L}w_{(n-1)L+m}c_{m-1}^{(l)}
\end{equation}
with zero mean and variance $2L\sigma_N^2T_{\rm b}/T_{\rm ch}=2L^2\sigma_N^2$. 

Note that if all beamforming gains were equal to $1$, \eqref{Eq:Decision_Stats_final} would become
\begin{equation}\label{Eq:Decision_Stats_no_gain}
\begin{split}
\hat{b}_n^{(l)} =& \sqrt{E_{\rm b}P(d)}\sum_{m=1}^{L}\left\{\left|\mathcal{L}_{(n-1)L+m,1}\right|\right.
\\&\hspace{2.02cm}\left.-\left|\mathcal{L}_{(n-1)L+m,2}\right|\right\}c_{m-1}^{(l)}+\bar{w}_n.
\end{split}
\end{equation}
But the number of transmitted ``$1$" chips minus the number of transmitted ``$-1$" chips at each chip period $T_{\rm ch}$ is equal to the sum of all transmitted chips, i$.$e$.$,
\begin{equation}\label{Eq:DSSS_no_gain}
\left|\mathcal{L}_{(n-1)L+m,1}\right| - \left|\mathcal{L}_{(n-1)L+m,2}\right| = \sum_{k=1}^{L} b_n^{(k)} c_{m}^{(k)}.
\end{equation}
Hence, by combining \eqref{Eq:Decision_Stats_no_gain}, \eqref{Eq:DSSS_no_gain} and \eqref{Eq:digital_despreading}, we can see that without beamforming gains the decision statistics for the $n$-th information bit of the $l$-th sensor node becomes
\begin{equation}\label{Eq:Decision_Stats_approx_no_gain}
\begin{split}
\hat{b}_n^{(l)} = L\sqrt{E_{\rm b}P(d)}b_n^{(l)} +\bar{w}_n.
\end{split}
\end{equation}

In the considered general case where there exist beamforming gains, it is not obvious why using \eqref{Eq:Decision_Stats_final} (direct bit level decoding) as the decision statistics for $\hat{b}_n^{(l)}$ is a good idea. Note that by taking into account the linearity of the average power gain illustrated in Fig$.$~\ref{gain_vs_num_nodes} we can write $\forall$ $q$:
\begin{equation}\label{Eq:3_2_Differences}
\sqrt{G\left(\mathcal{L}_{q,1}\right)}\left|\mathcal{L}_{q,1}\right| - \sqrt{G\left(\mathcal{L}_{q,2}\right)}\left|\mathcal{L}_{q,2}\right| \cong \left|\mathcal{L}_{q,1}\right|^\frac{3}{2} - \left|\mathcal{L}_{q,2}\right|^\frac{3}{2} 
\end{equation}
where the approximation includes the small deviation of the beamforming gain from linearity and its random fluctuation around the average for different sets of sensor nodes. Therefore, a more intuitive idea would be to first reverse the non-linearity by raising each received signal (including noise) to $2/3$ and then decode. Empirical results show however that the direct approach has a slightly better performance. This happens because raising the noisy signal to the $2/3$ introduces skewness to the random noise which turns out to be more harmful than the non-linearity of the received signal. In fact, a linear approximation can be employed based on the fact that (the proof is omitted due to space limitations) 
\begin{equation}\label{Eq:approx_error}
\frac{\left|\mathcal{L}_{q,1}\right|^\frac{3}{2} - \left|\mathcal{L}_{q,2}\right|^\frac{3}{2}
- \sqrt{L}\left(\left|\mathcal{L}_{q,1}\right| - \left|\mathcal{L}_{q,2}\right|\right)}{\left|\mathcal{L}_{q,1}\right|^\frac{3}{2} - \left|\mathcal{L}_{q,2}\right|^\frac{3}{2}} \leq 1-\frac{2\sqrt{2}}{3}.
\end{equation} 
Furthermore, the relative error is close to $1-\frac{2\sqrt{2}}{3} \cong 0.0572$ with very high probability so most of the time it holds: 
\begin{equation}\label{Eq:approx_error2}
\left|\mathcal{L}_{q,1}\right|^\frac{3}{2} - \left|\mathcal{L}_{q,2}\right|^\frac{3}{2}
\cong \frac{3}{2} \sqrt{\frac{L}{2}}\left(\left|\mathcal{L}_{q,1}\right| - \left|\mathcal{L}_{q,2}\right|\right).  
\end{equation} 
Thus, by employing the approximation of \eqref{Eq:approx_error2} into \eqref{Eq:Decision_Stats_final}, an approximate expression for $\hat{b}_n^{(l)}$ is obtained as 
\begin{equation}\label{Eq:Decision_Stats_approx}
\begin{split}
\hat{b}_n^{(l)} &\cong \sqrt{E_{\rm b}P(d)}\sum_{m=1}^{L}c_{m-1}^{(l)}\left\{\frac{3}{2}\sqrt{\frac{L}{2}}\left[\left|\mathcal{L}_{(n-1)L+m,1}\right|\right.\right.
\\&\hspace{2.45cm}\left.\left.-\left|\mathcal{L}_{(n-1)L+m,2}\right|\right]\right\}+\bar{w}_n 
\\&= \frac{3}{2}\sqrt{\frac{E_{\rm b}P(d)}{2}}L^{3/2}b_n^{(l)} +\bar{w}_n.
\end{split}
\end{equation}

\subsection{BER Analytical Approximation}
Let $\gamma_n^{(l)}$ be the SNR at RX for the $n$-th information bit of the $l$-th sensor node. Using the approximate expression for the decision statistics $\hat{b}_n^{(l)}$ given by \eqref{Eq:Decision_Stats_approx} and the variance of $\bar{w}_n$ given by \eqref{Eq:Sum_of_noise_terms}, a closed-form approximation for $\gamma_n^{(l)}$ can be easily obtained as
\begin{equation}\label{Eq:SNR}
\begin{split}
\tilde{\gamma}_n^{(l)} = \frac{9LE_{\rm b}P(d)}{16\sigma_N^2}  
\end{split}
\end{equation}
which is clearly independent of both indices $n$ and $l$. The latter expression for the SNR can be straightforwardly used to derive a closed-form approximate expression for the BER performance of the individual-data CB technique over an AWGN channel. In particular, using \cite[eq. (8.43)]{B:Sim_Alou_Book} for the error performance of BFSK modulation, the BER of the considered CB technique is given by 
\begin{equation}\label{Eq:Pe_awgn}
\tilde{P_{\rm e}} = Q\left(\frac{3\sqrt{LE_{\rm b}P(d)}}{4\sigma_N}\right)
\end{equation}
where $Q(\cdot)$ denotes the Gaussian $Q$-function \cite[eq. (4.1)]{B:Sim_Alou_Book}. 

Coming up with an exact analytical expression for $\gamma_n^{(l)}$ and BER performance would require analytical expressions for the probability distribution of RVs $G\left(\mathcal{L}_{t,i}\right)$'s (each depends on the respective $\left|\mathcal{L}_{t,i}\right|$), which seems to be a very complex problem. However, the BER approximation shown in \eqref{Eq:Pe_awgn} is very accurate as demonstrated in the simulation results presented in the following section.
\begin{figure}
\centering
\includegraphics[keepaspectratio,width=3in]{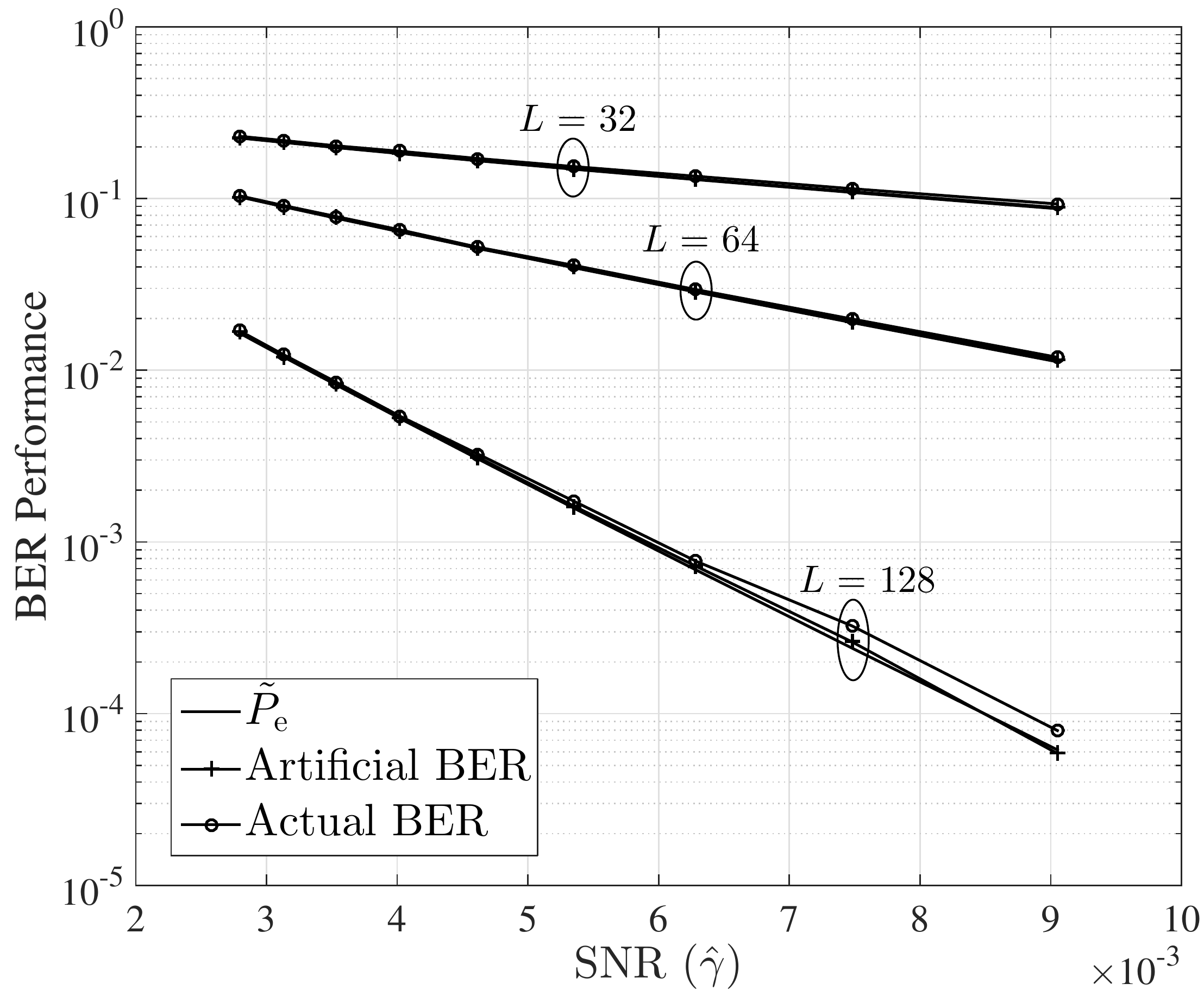}
\caption{BER performance of the individual-data CB technique as a function of the normalized SNR, $\hat{\gamma}$, for $L=32$, $64$ and $128$.}
\label{ber_vs_snr1}
\end{figure} 

\section{Performance Evaluation Results}\label{Sec:Results}
In order to validate the BER analysis of the previous section and compare the actual BER with the proposed analytical approximate expression given by \eqref{Eq:Pe_awgn}, we conducted the following \textsc{Matlab} simulation experiments. First, we placed sets of $L=64$, $32$ and $128$ sensor nodes at random positions uniformly distributed in a circular area with $R_{\max} = 100$ m. Then, for various values of the normalized SNR $\hat{\gamma} = E_{\rm b}P(d) / \sigma_N^2$ (this would be the SNR at RX without beamforming gain), each sensor node was assumed to transmit $10^4$ bits and the average BER over all $128\times10^4$ bits was calculated. As with the simulation results in Figs$.$~\ref{gain_vs_num_nodes} and~\ref{min_max_gain} in Section~\ref{Sec:Gain}, we have assumed $f_{1,2}=2.45$ GHz $\pm 64$ KHz. For each $\hat{\gamma}$ value three BER averages were calculated and depicted in Fig$.$~\ref{ber_vs_snr1}: \textit{i}) The actual BER obtained through Monte Carlo computer simulations where $A\left(\mathcal{L}_{q,i}\right)$ $\forall$ $q,i$ was explicitly calculated using \eqref{Eq:A_parameters} and \eqref{eq_directivity}; \textit{ii}) An artificial BER obtained by Monte Carlo computer simulations in which it was set $A\left(\mathcal{L}_{q,i}\right)=|\mathcal{L}_{q,i}|^{3/2}$ $\forall$ $q,i$; and \textit{iii}) The closed-form analytical BER approximation $\tilde{P_{\rm e}}$ given by \eqref{Eq:Pe_awgn}. 


The curves for the artificial BER and $\tilde{P_{\rm e}}$ almost coincide demonstrating that the approximation in \eqref{Eq:approx_error2} is very accurate. As for the actual BER performance, this is relatively close to the artificial BER, as the effect of the beamforming gain variability to the BER seems less significant than that of the AWGN. In fact, simulation results reveal that the BER caused by the effect of the beamforming gain variability alone (without any AWGN) is equal to zero. Furthermore, the quality (percentage of difference) of the approximation of the actual BER by the artificial BER is improving as $\hat{\gamma}$ decreases, since in this case the increasing AWGN noise plays a relatively bigger role to the resulting BER.

\section{Conclusions}
In this paper, we have analyzed the BER performance of an individual-data CB technique based on DSSS/BFSK which can be used to eliminate multi-hop communication in WSNs. The effect of varying beamforming gains per chip to the validity and BER performance of this technique has been analyzed and found to be relatively small. Furthermore, an analytical approximation of the BER of a bit level decoding scheme for the individual-data CB technique over an AWGN channel has been proposed and validated through simulation results. Having an expression for calculating the BER as a function of the number of transmitting nodes, transmit power, noise level and the distance to the common RX, allows for determining the required transmit power for a desired BER and thus energy budgeting and energy consumption evaluation of the individual-data CB technique. A comparison between the energy efficiency of this technique and other CB or multi-hop transmission techniques is outside of the scope of this paper and is left for future work.

\bibliographystyle{ieeetran}
\bibliography{reference_list}
\end{document}